\documentstyle[11pt,moriond02_kg_photo,epsfig]{article}

\def\be{\begin{equation}}
\def\ee{\end{equation}}
\def\bea{\begin{eqnarray}}
\def\eea{\end{eqnarray}}

\newcommand{\pom}{I\!\! P}
\begin{document}



%
\vspace*{3cm}
\title{MULTIGAP DIFFRACTION AT CDF
\footnote{
Presented at XXXVII$^{th}$ Rencontr\`es de Moriond, 
QCD and High Energy Hadronic Interactions,
Les Arcs, Savoie, France, March 16-23, 2002.}
}

\author{K. GOULIANOS}

\address{The Rockefeller University\\
1230 York Avenue, New York, NY 10021, USA\\
(For the CDF Collaboration)}

\maketitle\abstracts{
We present a study of $\bar pp$ collisions with a leading antiptoton and a 
rapidity gap in addition to that associated with the antiproton. The second gap
is either within the region available to the proton dissociation products, 
$\bar p+p\rightarrow (\bar p+gap)+X+gap+Y$, or adjacent to the outgoing proton
$\bar p+p\rightarrow (\bar p+gap)+X+(gap+p)$. 
Results are reported for two-gap to one-gap event 
ratios and compared with one-gap to no-gap ratios and with theoretical 
expectations. 
}
Diffractive $\bar pp$ interactions are characterized by a leading (high 
longitudinal momentum) outgoing proton or antiproton and/or a large  
{\em rapidity gap}, defined as a region of pseudorapidity, 
$\eta\equiv -\ln\tan\frac{\theta}{2}$, 
devoid of particles. The large rapidity gap is presumed to be due to the 
exchange of a Pomeron, which carries the internal quantum numbers of the 
vacuum. 
Rapidity gaps formed by multiplicity fluctuations in non-diffractive (ND) 
events are exponentially suppressed with $\Delta\eta$, so that gaps of 
$\Delta\eta>3$ are almost purely diffractive.
At high energies, where the available rapidity space is large, diffractive 
events may have more than one large gap. Using the Collider Detector at 
Fermilab (CDF), we have studied two types of events with two diffractive 
rapidity gaps in an event, shown schematically in Figs.~1 and 2.

\begin{minipage}[t]{7.5cm}
\vglue 0.1cm
\centerline{\psfig{figure=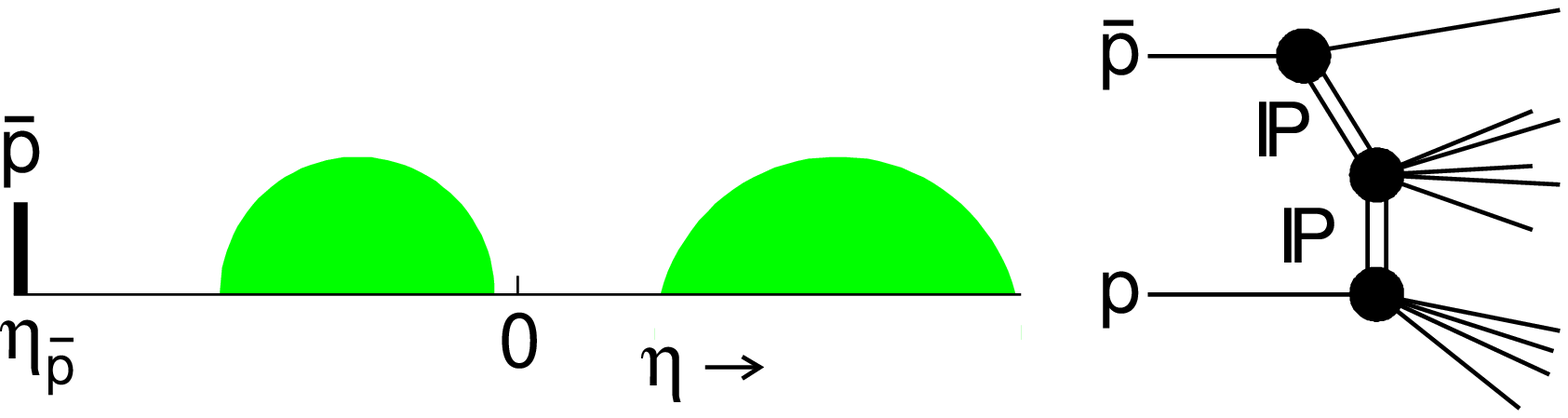,width=3in}}
\vglue -5.3cm
\centerline{Fig. 1: Single $\oplus$ Double Diffraction (SDD)}
\vglue -4in
\end{minipage}
\begin{minipage}[t]{7.5cm}
\vglue -0.2cm
\centerline{\psfig{figure=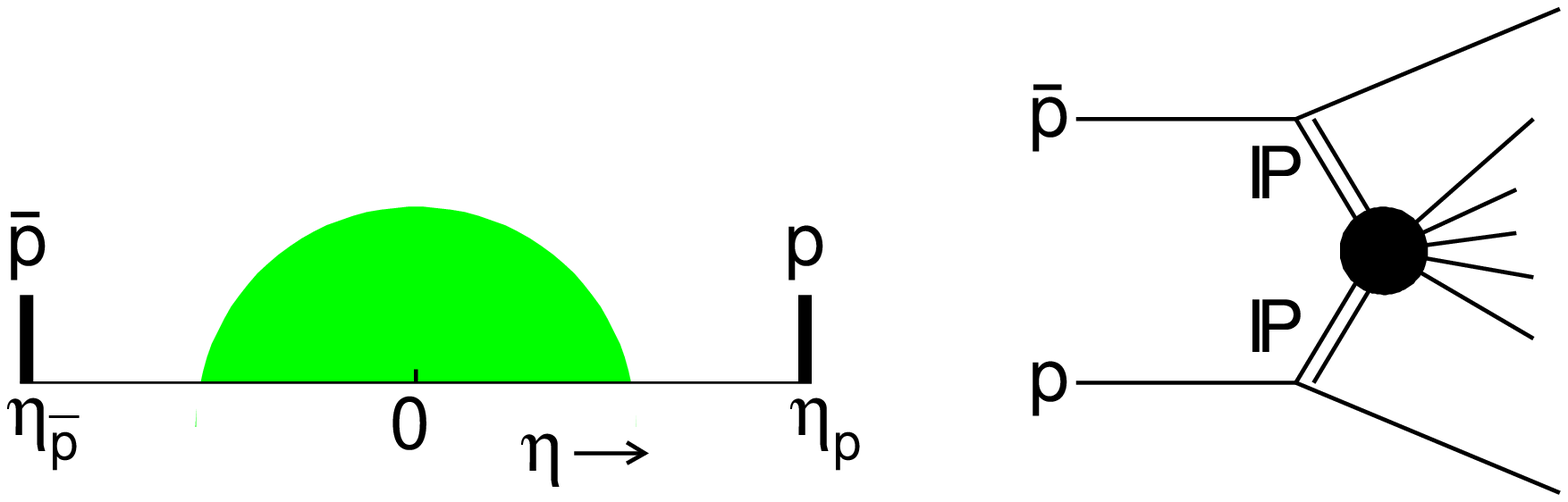,width=3in}}
\vglue -5cm
\centerline{Fig. 2: Double Pomeron Exchange (DPE)}
\vglue -4in
\end{minipage}

\newpage
The motivation for this study is its potential for providing further
understanding of the underlying mechanism responsible for 
the suppression of diffractive cross sections at high energies relative 
to Regge theory predictions. As shown in Figs. 3 and 4, such a suppression 
has been observed for both single diffraction (SD), 
$\bar p(p)+p\rightarrow [\bar p(p)+gap]+X$, and double diffraction (DD),
$\bar p(p)+p\rightarrow X_1+gap+X_2$.

\noindent\begin{minipage}[t]{3in}
\vspace*{-1.4cm}
\centerline{\hspace*{-0.1cm}
\psfig{figure=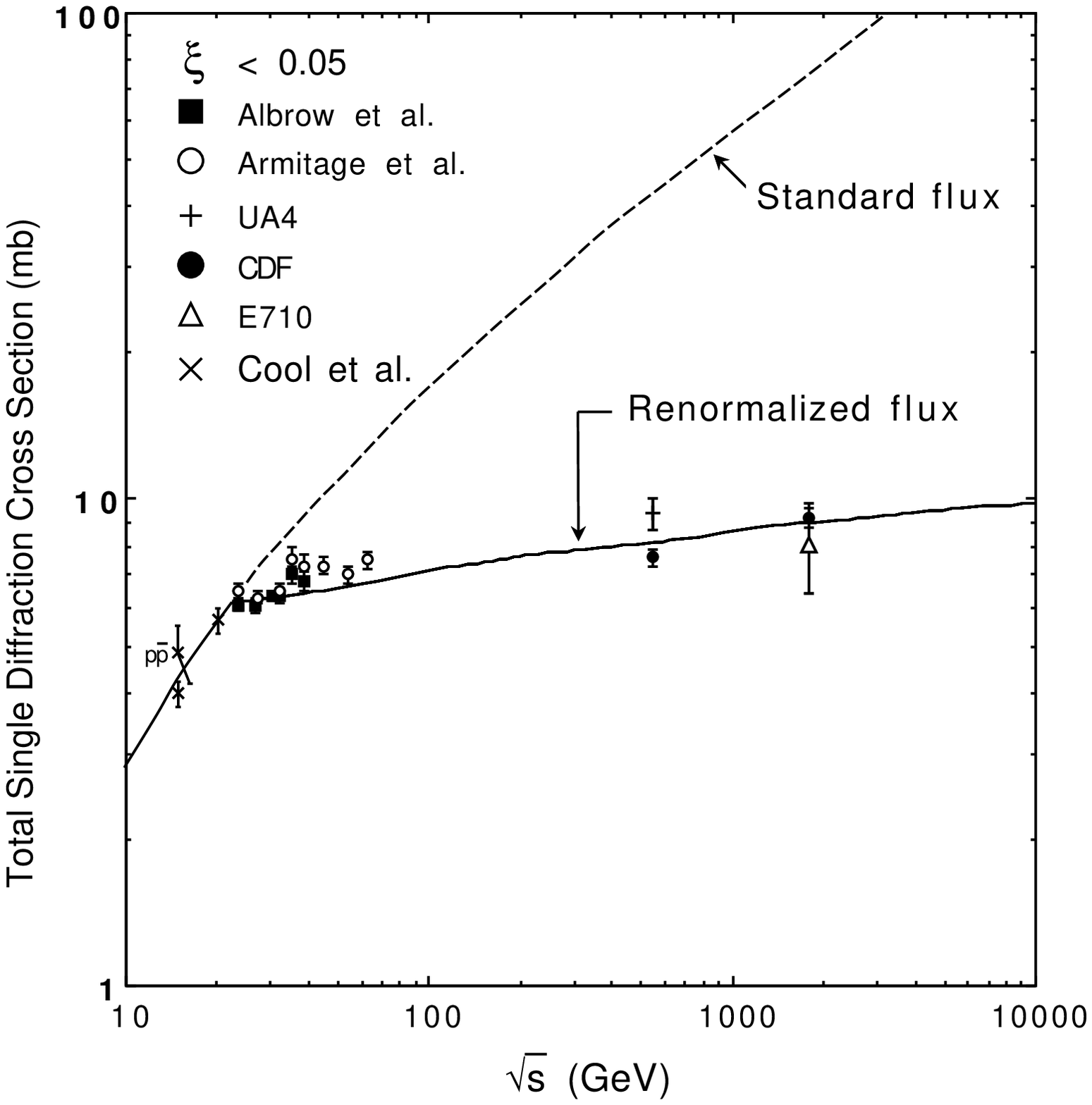,width=3.35in}}
\vglue -2.55cm
FIG. 3: The $\bar pp$ total SD
cross section exhibits an $s$-dependence consistent
with the renormalization procedure of Ref.~\cite{R},
contrary to
the $s^{2\epsilon}$ behaviour expected from Regge theory
(figure from Ref.~\cite{R}).
\end{minipage}
\hspace*{0.2in}
\begin{minipage}[t]{3in}
\vspace*{-0.4cm}
\centerline{\psfig{figure=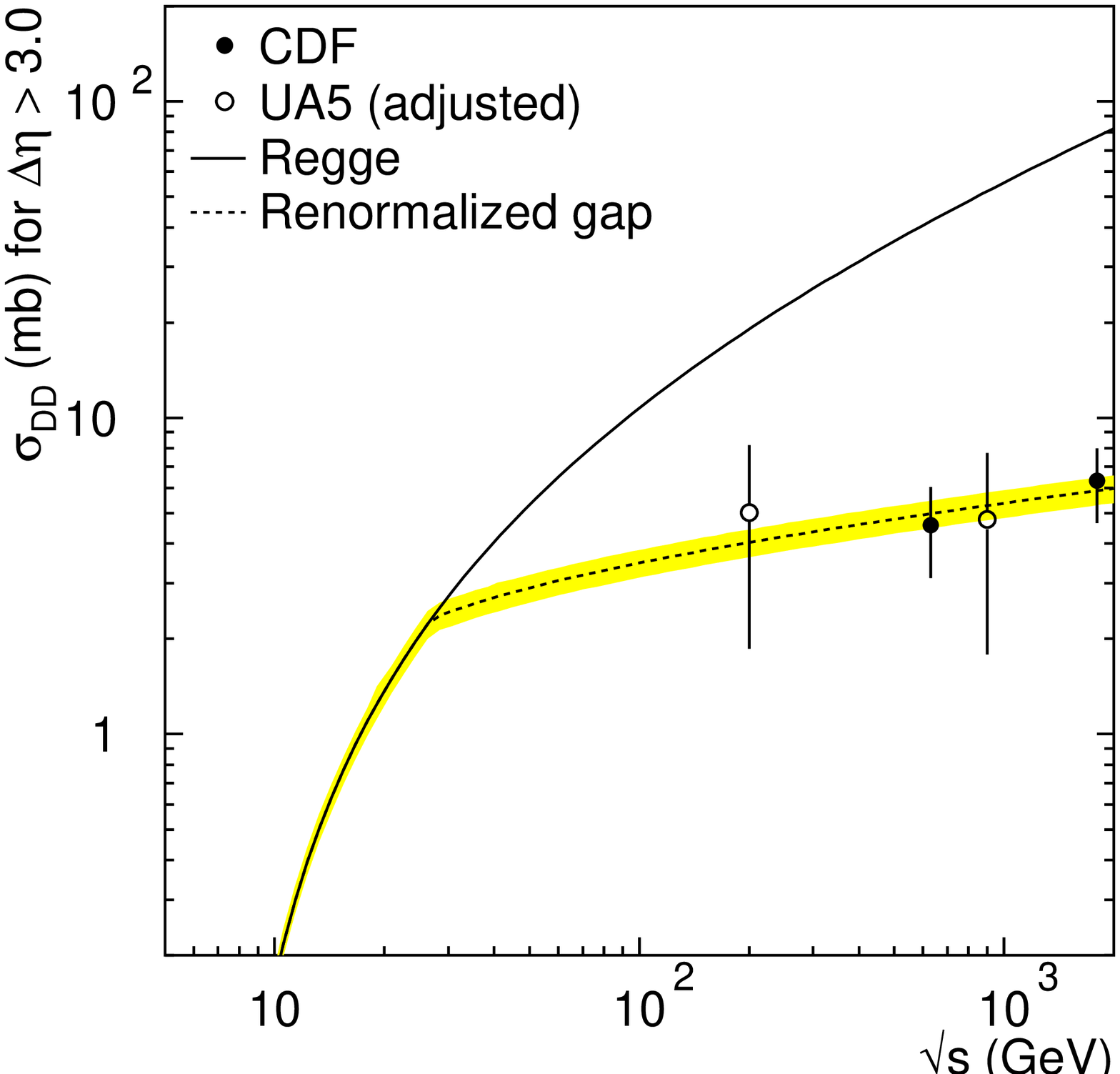,width=3in}}
\vglue 0.05cm
FIG. 4: The $\bar pp$ total DD (central gap) cross section
agrees with the prediction of the
{\em renormalized rapidity gap} model~\cite{KGgap},
contrary to the $s^{2\epsilon}$
expectation from Regge theory (figure from Ref.~\cite{dd}).
\end{minipage}

Naively, the suppression relative to Regge based predictions is
attributed to the spoiling of the diffractive 
rapidity gap by color exchanges in addition to Pomeron exchange.
In an event with two rapidity gaps, additional color exchanges would 
generally spoil both gaps. Hence, ratios of two-gap to one-gap 
rates should be unsuppressed. Measurements of such ratios could 
therefore be used to test the QCD aspects of gap formation without 
the complications arising from the rapidity gap survival probability.
 
The data used for this study are inclusive SD event samples at $\sqrt s=1800$
and 630 GeV collected by triggering on a leading antiproton detected in 
a Roman Pot Spectrometer (RPS)~\cite{jj1800,jj630}. Below, we list the 
number of events used in each analysis within the indicated regions of 
antiproton fractional momentum loss $\xi_{\bar p}$ and 
4-momentum transfer squared $t$, 
after applying the vertex cuts $|z_{vtx}|<60$ cm and 
$N_{vtx}\le 1$ and a 4-momentum squared cut of $|t|<0.02$ GeV$^2$ (except for 
DPE at 1800 GeV for which $|t|<1.0$ GeV$^2$):

\begin{center}
\begin{tabular}{lccc}
Process&$\xi$&Events at 1800 GeV&Events at 630 GeV\\
\hline\hline
SDD&$0.06<\xi<0.09$&412K&162K\\
DPE&$0.035<\xi<0.095$&746K&136K\\
\hline
\end{tabular}
\end{center}

In the SDD analysis, the mean value of $\xi=0.07$ corresponds to a 
diffractive mass of $\approx 480\;(170)$ GeV at $\sqrt s=1800$ (630) GeV. 
The diffractive cluster X in such events covers almost the entire 
CDF calorimetry, which extends through the region $|\eta|<4.2$. 
Therefore, we use the same method of analysis as that used to extract the gap 
fraction in the case of DD~\cite{dd}. We search for {\em experimental gaps}
overlapping $\eta=0$, defined as regions of $\eta$ with no tracks or calorimeter
towers above thresholds chosen to minimize calorimeter noise contributions.
The results, corrected for triggering efficiency of $BBC_p$ (the beam 
counter array on the proton side) and converted to {\em nominal gaps} 
defined by $\Delta\eta=\ln\frac{s}{M_1^2M_2^2}$, are shown in Figs. 5 and 6.

\noindent\begin{minipage}[t]{3in}
\vspace*{0.1cm}
\psfig{figure=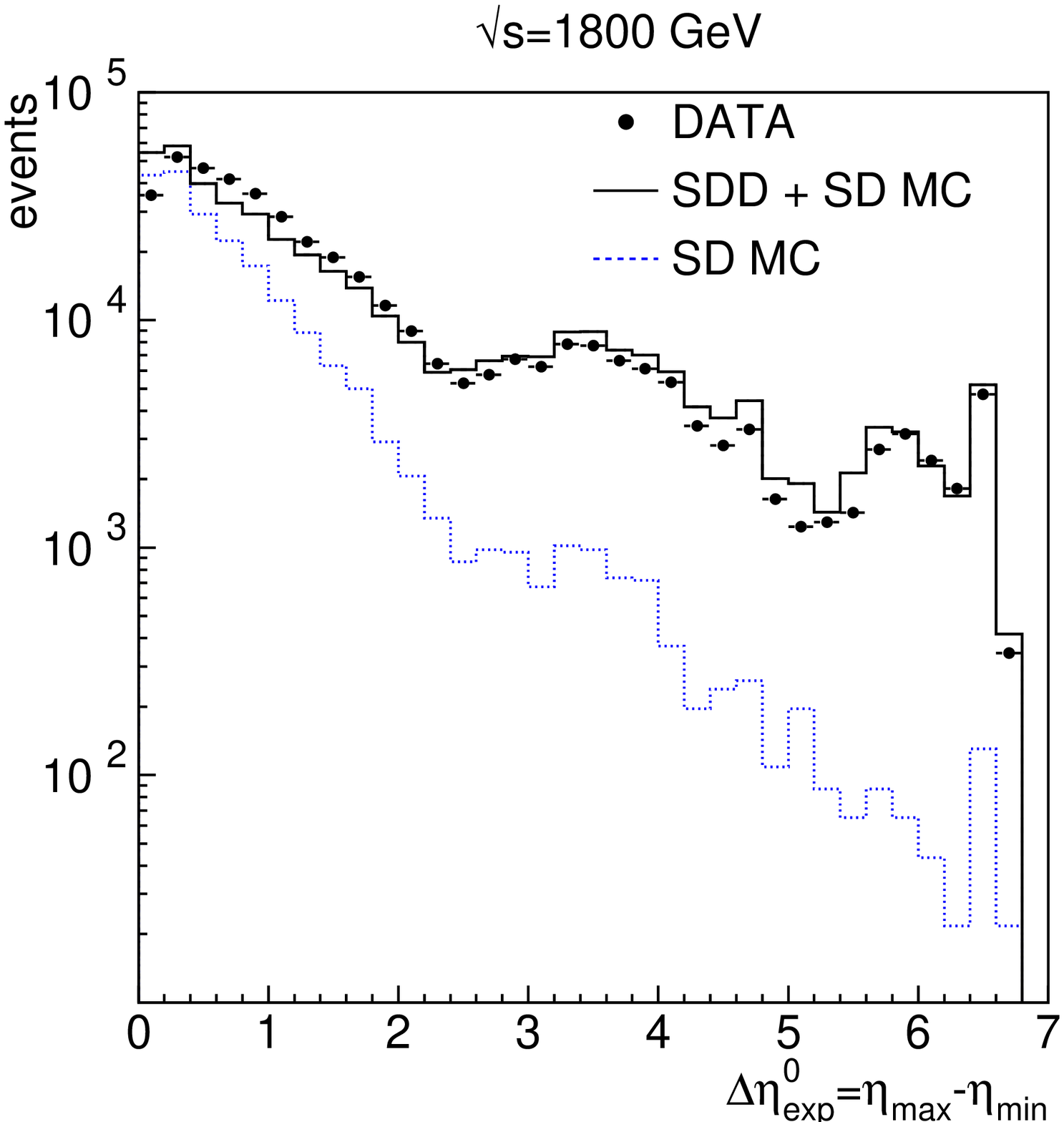,width=3in}
\vspace*{0.2cm}
Fig. 5: The number of events as a function of
$\Delta\eta_{exp}^0=\eta_{max}-\eta_{min}$ for data at
$\protect\sqrt s=1800$ GeV (points), for SDD  Monte Carlo generated events
(solid line), and for only SD Monte Carlo events (dashed line).
\end{minipage}
\hspace*{0.2in}
\begin{minipage}[t]{3in}
\vspace*{0.4cm}
\centerline{\hspace*{0.5cm}\psfig{figure=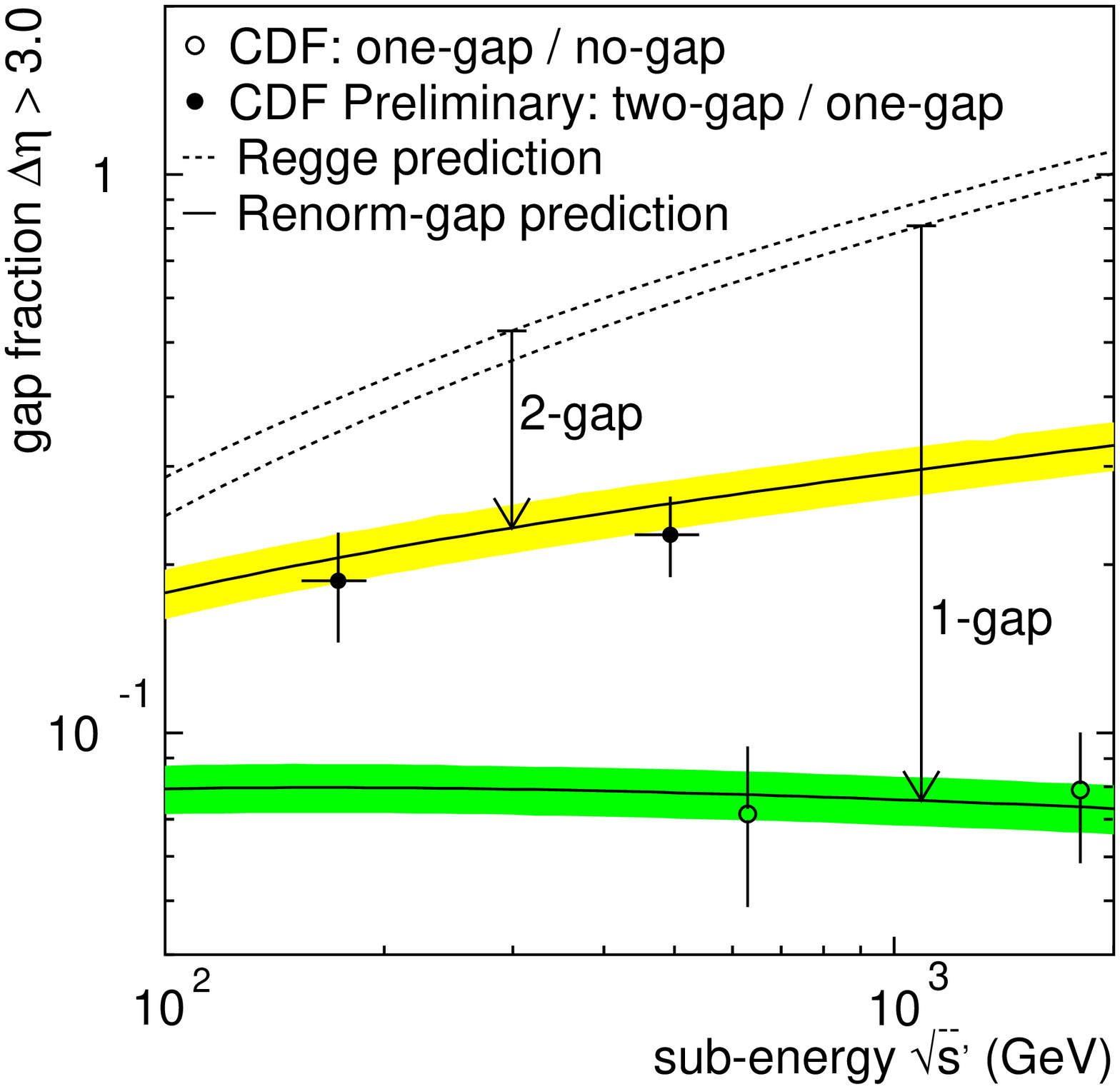,width=3in}}
\vspace*{-0.1cm}
FIG. 6: Ratios of SDD to SD rates (points)
and DD to total (no-gap) rates (open circles) as a function
of $\sqrt{s'}$ of the sub-process $\pom p$  and of $\bar pp$,
respectively.
The uncertainties are highly correlated among all data points.
\end{minipage}
\vglue 1ex
The SDD Monte carlo simulation is based on Regge theory Pomeron exchange 
with the normalization left free to be determined from the data. 
The differential $dN/d\Delta\eta^0$ shape agrees with the theory (Fig.~5), but 
the two-gap to one-gap ratio is suppressed (Fig.~6). However, the 
suppression is not  as large that as in the one-gap to no-gap ratio. The bands 
through the data points represent predictions of the renormalized multigap 
parton model approach to diffraction~\cite{multigap}, 
which is a generalization of the renormalization models 
used for single~\cite{R} and double~\cite{KGgap} diffraction.   

In the DPE analysis, the $\xi_p$ is measured from calorimeter and 
beam counter information using the formula below and summing over 
all particles, defined experimentally as 
beam-beam counter (BBC) hits or calorimeter towers above 
$\eta$-dependent thresholds 
chosen to minimize noise contributions.
$$\xi^{\rm\sc X}_p=\frac{M^2_{\rm\sc X}}{\xi_{\bar p}\cdot s}=
\frac{\sum_i E_{\rm\sc T}^i\;\exp(+\eta^i)}{\sqrt s}$$
\noindent For BBC hits we use the average value of $\eta$ of the BBC segment 
of the hit and an $E_T$ value 
randomly chosen from the expected $E_T$ distribution. 
The $\xi^X$ obtained by this method was calibrated by comparing 
$\xi^X_{\bar p}$, obtained by using $\exp(-\eta^i)$ in the above equation,
with the value of $\xi_{\bar p}^{RPS}$ measured by the Roman Pot Spectrometer.

Figure 7 shows the $\xi^X_{\bar p}$ distribution for $\sqrt s=$1800 GeV.
The bump at $\xi_{\bar p}^X\sim 10^{-3}$ is attributed to
central calorimeter noise and is reproduced in Monte Carlo simulations. 
The variation of tower $E_T$ threshold across the various 
components of the CDF calorimetry does 
not affect appreciably the slope of the $\xi^X_{\bar p}$ distribution. 
The solid line represents the distribution measured in SD~\cite{PRD}.
The shapes of the DPE and SD distributions are in good agreement all the way 
down to the lowest values kinematically allowed.  

\noindent\begin{minipage}[t]{3in}
\vspace*{1.1cm}
\centerline{\psfig{figure=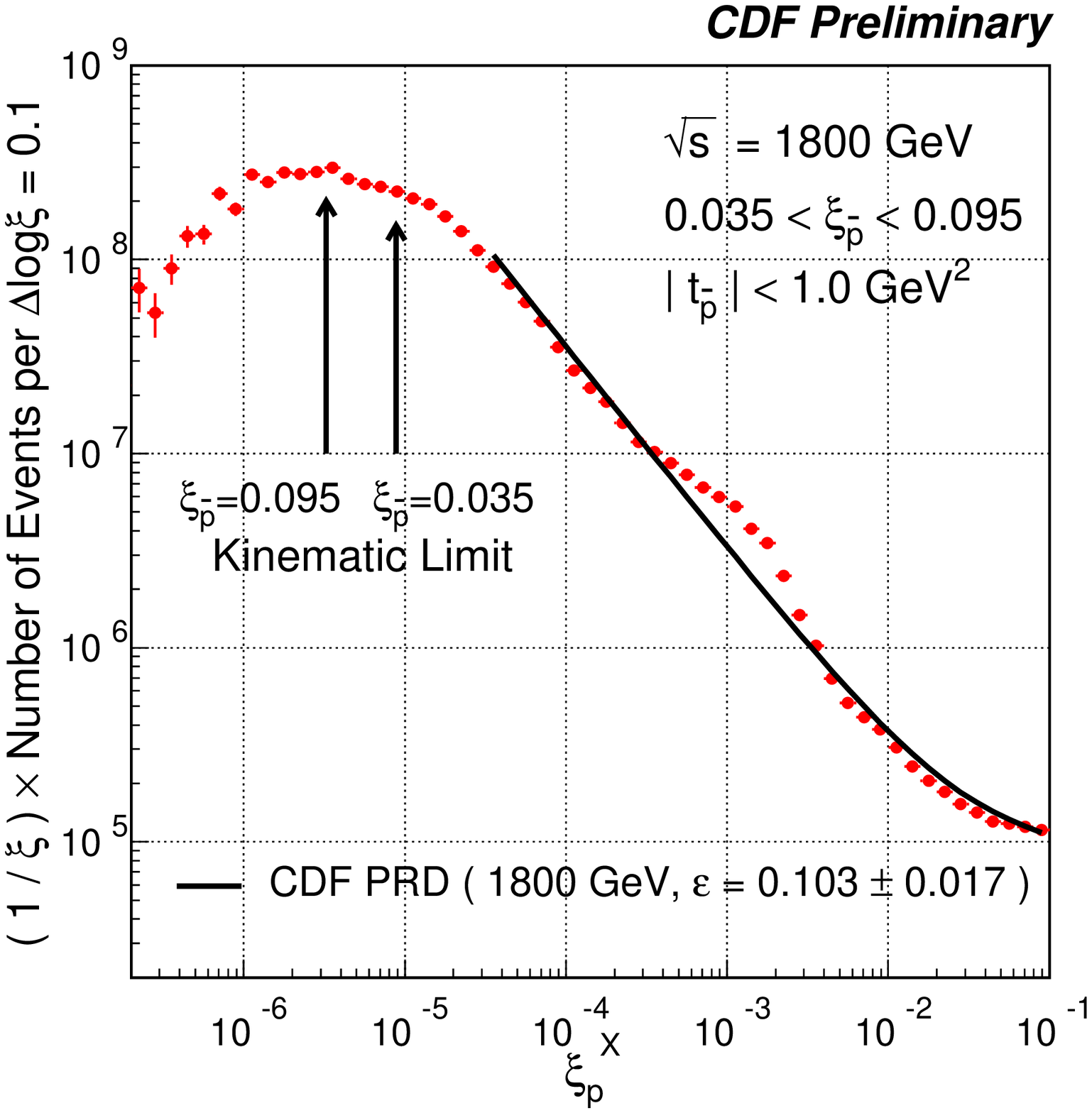,width=3in}}
\vspace*{-2cm}
Fig. 7: $\xi_{\bar p}^X$ distribution at $\sqrt s=$1800 GeV 
for events with a $\bar p$ of 
$0.035<\xi_{\bar p}^{RPS}<0.095$. 
The solid line is the distribution obtained in single
diffraction dissociation. The bump at $\xi_{\bar p}^X\sim 10^{-3}$ is due to 
central calorimeter noise and is reproduced in Monte Carlo simulations.
\end{minipage}
\hspace*{0.2in}
\begin{minipage}[t]{3in}
\vspace*{1.1cm}
\centerline{\psfig{figure=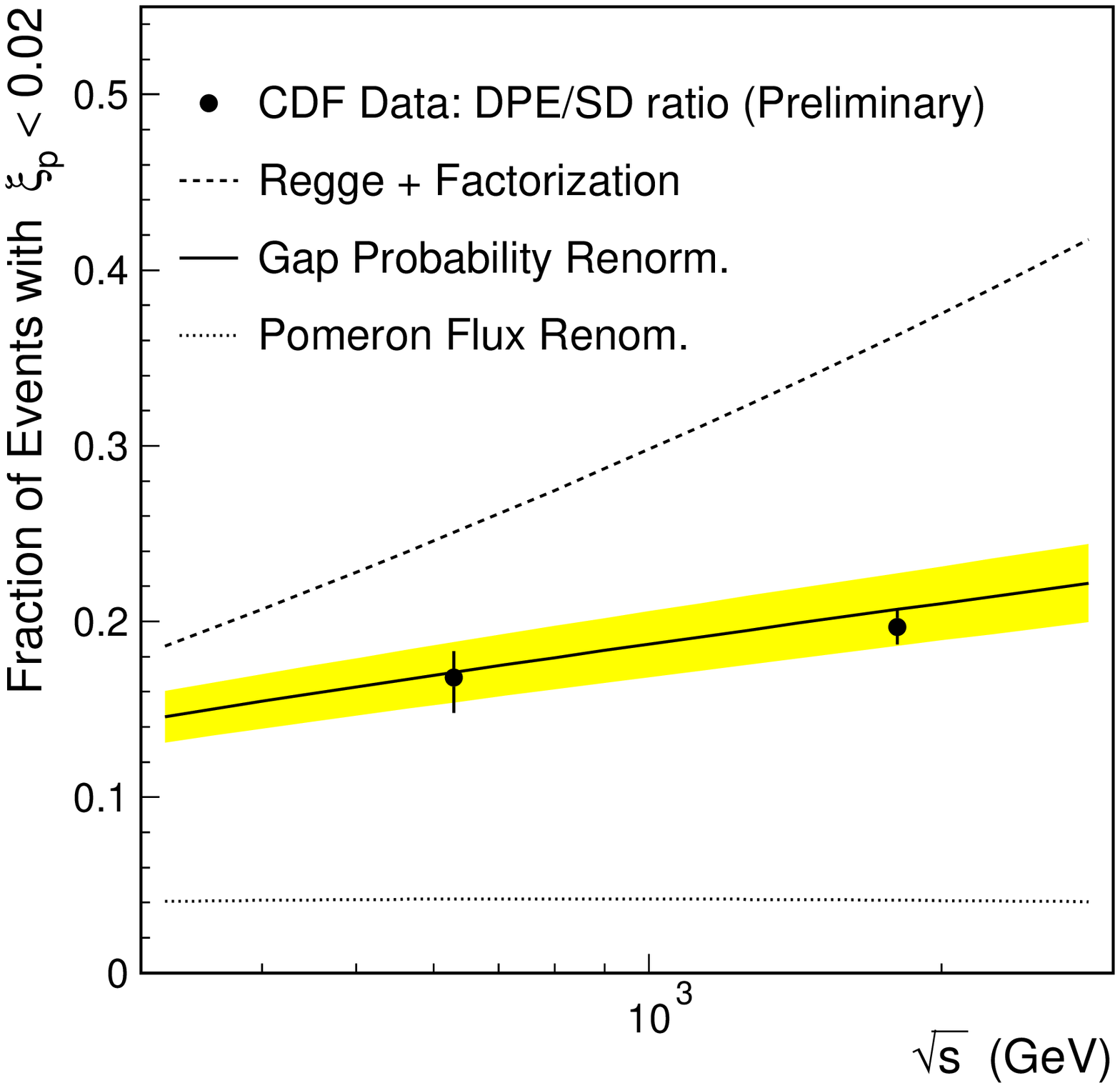,width=3in}}
\vspace*{-2cm}
FIG. 8: Measured ratios of DPE to SD rates (points)
compared with predictions based on Regge theory(dashed), Pomeron flux 
renormalization for both exchanged Pomerons (dotted) and gap probability 
renormalization (solid line).
\end{minipage}
\vglue 1ex
The ratio of DPE to inclusive SD events was evaluated for $\xi_p^X<0.02$.
The results for $\sqrt s=$1800 and 630 GeV are presented in the table below 
and in Fig.~8. Also presented are the 
expectations from gap probability renormalization~\cite{multigap}, 
Regge theory and factorization, and Pomeron flux renormalization 
for both exchanged Pomerons~\cite{R}. The  quoted uncertainties
are largely systematic for both data and theory; the theoretical 
uncertainties of 10\% are due to the uncertainty in the ratio of the 
triple-Pomeron to the Pomeron-nucleon couplings~\cite{GM}. 
The data are in excellent agreement with the predictions of the gap 
renormalization approach.
\vspace*{0.3cm}
\begin{center}
\begin{tabular}{lcc}
Source&$R^{DPE}_{SD}$(1800 GeV)&$R^{DPE}_{SD}$(630 GeV)\\
\hline\hline
Data&$0.197\pm 0.010$&$0.168\pm 0.018$\\
$P_{gap}$ renormalization&$0.21\;\pm0.02$&$0.17\;\pm0.02$\\
Regge $\oplus$ factorization&$0.36\;\pm 0.04$&$0.25\;\pm 0.03$\\
$\pom$-flux renormalization&$0.041\pm 0.004$&$0.041\pm 0.004$\\
\hline
\end{tabular}
\end{center}

\end{document}